# REVIEW

## Advances and Applications of Dynamic Surface-Enhanced Raman Spectroscopy (SERS) for Single Molecule Studies

Yanqiu Zou,[a] Huaizhou Jin,[b] Qifei Ma,[c] Zhenrong Zheng,*[a] Shukun Weng,[d] Karol Kolataj,[e] Guillermo Acuna,[e] Ilko Bald[f] and Denis Garoli*[c,d,g]

Dynamic surface-enhanced Raman spectroscopy (SERS) is nowadays one of the most interesting applications of SERS, in particular for single molecule studies. In fact, it enables the study of real-time processes at the molecular level. This review summarizes the latest developments in dynamic SERS techniques and their applications, focusing on new instrumentation, data analysis methods, temporal resolution and sensitivity improvements, and novel substrates. We highlight the progress and applications of single-molecule dynamic SERS in monitoring chemical reactions, catalysis, biomolecular interactions, conformational dynamics, and real-time sensing and detection. We aim to provide a comprehensive review on its advancements, applications as well as its current challenges and development frontiers.

## Introduction

### Introduction to Dynamic SERS

Raman Spectroscopy (RS) is a powerful analytical technique in various fields, including chemistry, biology, and materials sciences(1, 2). Compared to traditional detection and sensing methods in these sciences such as mass spectroscopy and High-Performance Liquid Chromatography (HPLC) in chemistry and material sciences or enzyme-linked immunosorbent assay (ELISA) in biology, RS provides rich structural information, or "vibrational fingerprints" that can be used for identification and characterization. Also compared to infrared spectroscopy, RS is suitable for the analysis of aqueous samples, because water has a strong absorption in the IR region that can obscure IR signals, whereas the Raman scattering from water is very weak.

Despite the inherent advantages of RS, its practical application has been limited by its relatively low sensitivity. This limitation has been addressed by the development of SERS(3-10), which uses electromagnetic fields generated by plasmonic nanostructures to amplify the Raman scattering signal of molecules. The SERS enhancement factor can reach up to $10^{12}$, allowing for the detection of single molecules(11-15). Recent advancements in the design and fabrication of SERS substrates have significantly contributed to the improved sensitivity and reproducibility of SERS measurements(16-19). Novel plasmonic nanostructure have been developed to provide high-density hotspots for enhanced SERS signals(20, 21).

In recent years, the development of dynamic SERS has garnered significant attention, as it enables real-time monitoring of chemical and biological processes. Dynamic SERS can be used to provide insights into the kinetics and mechanisms of these processes(22, 23). Dynamic SERS involves decreasing the acquisition time from minutes or seconds down to milliseconds to effectively capture, with high temporal resolution, SERS signals during potential chemical or mechanical reactions of the molecules under investigation. The fundamental concept behind dynamic SERS exploits the dynamic nature of nanoparticles in solution to extract pure SERS signals from the overwhelming solvent background. Through analysis of the spectral fluctuations, dynamic SERS effectively removes spectral interferences and enables the detection of SERS signals from nanoparticles at extremely low concentrations, which shows both in the Figure 1(a), (b). The key to the development of dynamic SERS is the development of both sensitivity and temporal resolution. High SERS sensitivity allows to detect trace amounts of molecules at very little time intervals, thus enabling researchers to continuously monitor the molecular structures of the analytes and discover molecular confirmations or reaction intermediates that appear for very short time. Up to date, some research groups were able to collect Raman spectra in the consecutive mode at the millisecond or microsecond intervals(24, 25) (Figure 1(d)). The ability to detect single molecules has also opened up new avenues for dynamic SERS, such as studying conformations and intermediates by observe and track single molecules in real-time(26).

The growing interest in dynamic SERS is demonstrated from the steady increase in the number of publications on this topic over the years. As shown in Figure 1(d), dynamic SERS first became a topic around 2008. Since then, the annual number of publications has steadily grown from below 15 per year before 2012 to more than 50 publications per year after 2017, peaking at 86 publications in 2022. This trend highlights the increasing recognition of dynamic SERS as a promising tool for investigating complex chemical and biological systems, as the sensitivity and temporal resolution of dynamic SERS continues to evolve.

### Key foundational works

The development of dynamic SERS is based on a series of key foundational studies. These pioneering works first established the existence of transient and amplified Raman signal, and improved on both sensitivity and temporal resolution.

One of the earliest demonstrations of dynamic SERS in a biologically relevant system was the study by Etchegoin et al. on photoinduced oxygen release in hemoglobin(27). By means surface-enhanced resonant Raman scattering (SERRS) using silver colloids, they observed transient, massive amplifications of Raman signals during oxygen dissociation, attributed to a charge-transfer mechanism at the protein-nanoparticle interface that allows normally Raman-forbidden modes to be observed. The enhanced signals were shown to originate from the heme groups and be sensitive to their local environment and orientation. Yan et al. studied optimal hotspot generation for dynamic SERS, and monitored the formation of optimal hotspots in real-time using in-situ small-angle X-ray scattering(28).

dynamic control and sub-diffraction localization of plasmonic

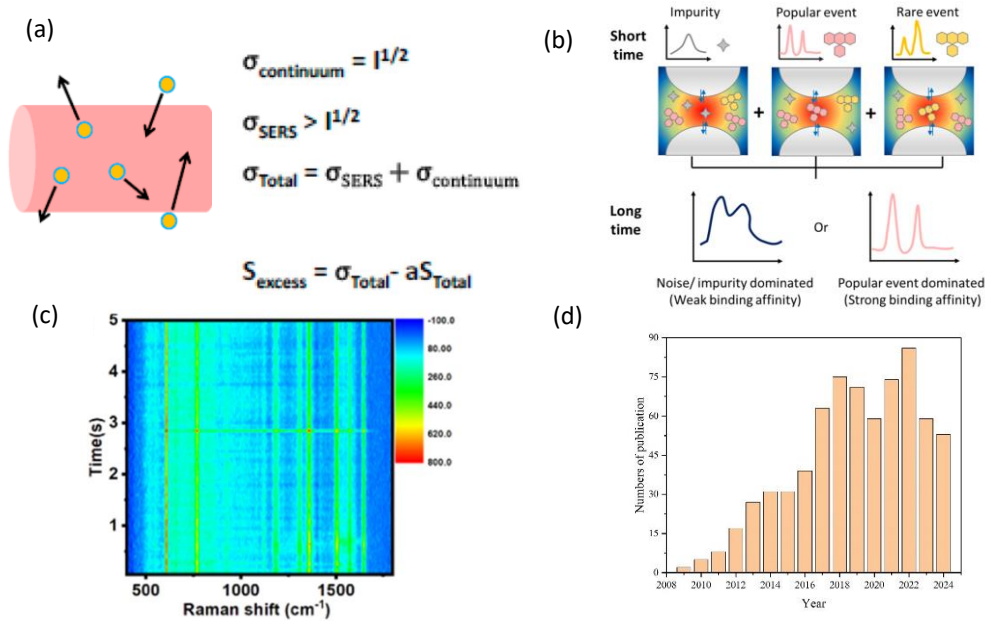

Figure 1 Dynamic SERS concept and its research trends. (a) (Left) An illustration of dynamic SERS concept which shows a colloidal nanoparticle moving through a focused laser beam. (Right) The standard deviation of the continuum, $\sigma_{continuum}$, will scale as the square root of the intensity while the $\sigma_{SERS}$ from the nanoparticle will be larger. The signal attributed to the excess noise, $S_{excess}$, from the dynamic SERS-active nanoparticles can be determined by calculating the difference between the total noise, $\sigma_{total}$, and the total signal, $S_{total}$, while accounting for the disparity in magnitude between the standard deviation and average signals using a factor a. Reproduced with permission from ref (25). Copyright 2012 American Chemical Society. (b)This updated schematic illustrate rare event to single molecule level. In the traditional SERS measurement (long time exposure), either the popular events or the noise-dominated SERS spectrum is obtained. In the dynamic SERS measurement (short time exposure), the signals from the impurity and popular and rare events of the target are well distinguished; (c)Time-dependent heat map of SERS intensity as a function of Raman shift in R6G solution. Reproduced with permission from ref (24). Copyright 2020 American Chemical Society. (d)Numbers of literatures published annually since 2024, searched for using the keyword "dynamic SERS" by using Google Scholar, as of August 2024.

The temporal resolution of dynamic SERS slowly improved. In 2012, Scott et al. were able to acquire a clear Raman spectrum of 4-mercaptopyridine (4MP) every 100ms by exploiting nanoparticle Brownian motion to extract site-specific SERS signals and remove interfering background contributions(25). They were able to acquire 1000 consistent SERS spectra over the span of 10 seconds.

Dynamic SERS mapping and imaging is another research frontier for dynamic SERS. In 2014, Ertsgaard et al. demonstrate hotspots on a silver nanohole array, which enabled super-resolution SERS imaging (29). The work of Brulé et al. is one of the earlier works for data analysis in dynamic SERS. They used statistical and Fourier analysis of Raman intensity fluctuations to determine the concentration sensitivity of SM dynamic SERS(30). Their probability density function allows calibration-free determination of molecular concentrations over a wide $10^{-11}$ to $10^{-6}$ M range.

**Purpose and scope of the review**

The purpose of this review is to provide a comprehensive overview of the recent advances in single molecule dynamic SERS, its mechanisms and applications, as well as methods or strategies that can be used together with dynamic SERS. The review aims to include these primary applications of dynamic SERS in research, conformational dynamics of biomolecules, nucleic acid and protein sequencing, chemical reaction dynamics and dynamic SERS imaging.

# Recent advancements in dynamic SERS techniques

## Advancements in fast dynamic SERS

High-speed imaging and spectroscopy techniques have played a crucial role in advancing dynamic SERS by enabling the acquisition of SERS spectra with improved temporal resolution. The initial connection between fluctuations in SERS intensity and the measurable response from single molecules has been a significant focus in the field of SERS(5, 9). This relationship highlights the sensitivity of SERS, which allows for the detection of individual molecules through enhanced Raman signals, particularly at specific sites on metallic surfaces known as SERS active sites or hot spots. Most of the work done on this field are by Linquist and Brolo's group. In 2019 they developed a high-speed imaging system that combines a fast camera and a high-

In 2023, the same group achieved ultrafast dynamic SERS measurement, collecting 100 thousand spectra per second and pushed temporal resolution to 10 μs(34). Ultimately, Cheng et al. pushed the limit of temporal resolution to 5 μs measuring 200 thousand spectra per second using a lab-built 32-channel tuned-amplifier array(35). Employing this measuring system in stimulated Raman scattering experiments the authors were able to obtain 200 thousand spectra, or measure 11,000 individual particles per second using a flow cytometry set-up. Their works on ultra-high-speed SIFs open new potential applications in analytical chemistry and chemical imaging. Understanding plasmon-induced fluctuations can enable ultra-power laser to capture SERS intensity fluctuations (SIFs) from individual nanoparticles at a rate of up to 800,000 frames per second(31). Details of their experimental approach also with the high-speed SERS signal acquisition are showed in Figure 2. This technique allowed them to observe the formation and dissociation of single-molecule SERS hotspots in real-time, providing insights into the dynamic nature of SERS-active sites.

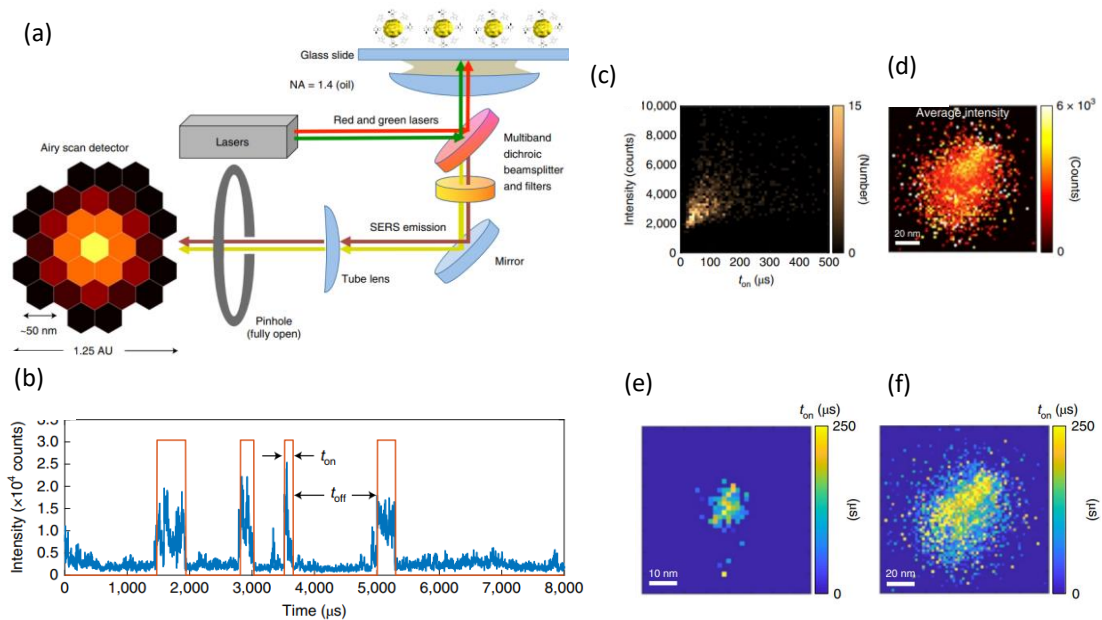

Figure 2 Classical high-speed imaging SERS system and single molecule SIFs. (a) Schematic illustrations of High-speed spectral acquisition system and their experimental approach. Single-molecule SIFs, covered 500 cm$^{-1}$ to 2000 cm$^{-1}$, (b) digitized (d-SIF, shown as a red envelope) to extract time dynamics. (c) Bivariate histogram of the same data showing the SIF intensity versus the d-SIF on time $t_{on}$ in microseconds. (d) An image of another nanoparticle, this time representing 11,577 individual SIF events, and again showing the occurrence of SIFs all over the single particle. (e) The 32 pixels Airyscan imaging array was used to spatially map the characteristics of the nanoparticle from SM-SERS events. (f) Spatial distribution of $t_{on}$ for the same particle as in (d). Reproduced with permission from ref (31). Copyright 2019 Springer Nature.

Building upon this work, they further investigated the ultra-high-speed dynamics of SERS hotspots, improving the temporal resolution to 100 μs in 2020(32). This advancement allowed them to probe the sub-millisecond dynamics of SERS hotspots, revealing previously unobserved chemical and biological phenomena, while also enabling to research more on the phenomena of SERS intensity fluctuations (SIFs) itself(33). They found that high-speed SIF events occur with relatively equal probability across a broad spectral range, including both the Stokes and anti-Stokes regions of the SERS spectrum.

low concentration quantification, while recording fluctuation frequency changes may mitigate secondary effects. Super-resolution techniques, such as STORM, can be adapted to utilize SIFs for imaging dynamic events in single cells, driving further discoveries in SERS-based imaging.

De Albuquerque et al. also used SIFs for dynamic SERS research. They integrated high-speed SERS imaging with super-resolution microscopy to visualize the spatial distribution and temporal evolution of SERS hotspots on single nanoparticles(36, 37). This multi-modal approach allowed them to correlate the SERS intensity fluctuations with the nanoscale morphological changes of the nanoparticles, providing a deeper insight into the structure-function relationships in SERS.

**Dynamic SERS related imaging techniques**

The idea of dynamic SERS imaging first appeared around 2009 to 2010. In 2009, Bozzini et al used dynamic SERS imaging to study the electrodeposition of gold (Au) from a potassium gold cyanide bath, and were able to analyse the time-dependent SERS intensities of adsorbed $CN^-$(38). In 2010, Wark et al. introduce a wide-field imaging approach for the real-time analysis of SERS-active silver nanoparticle clusters suspended in solution(39). They used a 550 nm bandpass filter to select the 1275 $cm^{-1}$ Raman band of the ABT-DMOP reporter molecule and were able to take the image of the peak over time, doing so, they were able to collect sufficient signals from an integration time of 50 ms.

Fast forward to the recent development of dynamic SERS imaging techniques, Linquist et al. utilized high speed imaging to study single-molecule dynamic SERS intensity fluctuations(40). High-speed imaging was used to reveal hotspot-to-hotspot variability, suggesting that transient atomic-scale hotspot generation dominates over molecular diffusion. Dynamic SERS with ms time resolution was also used to detect nanometer-scale disorder in Au crystal lattices by detection of plasmonic flare emissions due to the formation of plasmonic pico-cavities(41).

These techniques are now utilized in various application in chemistry, cell biology and so on. In cell biology, dynamic SERS can be used to achieve sensitive and multiplexed detection of drugs, intracellular pH, and reactive oxygen species in living cells with improved spatiotemporal resolution(42). For example, Koike et al. utilized 3D live-cell SERS imaging to visualize the drug uptake process with sufficient temporal resolution(43). By introducing gold nanoparticles into lysosomes and using an alkyne-tagged cathepsin B inhibitor, they achieve sensitive detection and 3D time-lapse imaging of drug localization.

**Novel substrates and platforms**

There are several types of new SERS substrates and platforms being developed for dynamic SERS. Nanorods(44), nanocages(45), nanopores(46, 47) or nanoslits(48) are all novel

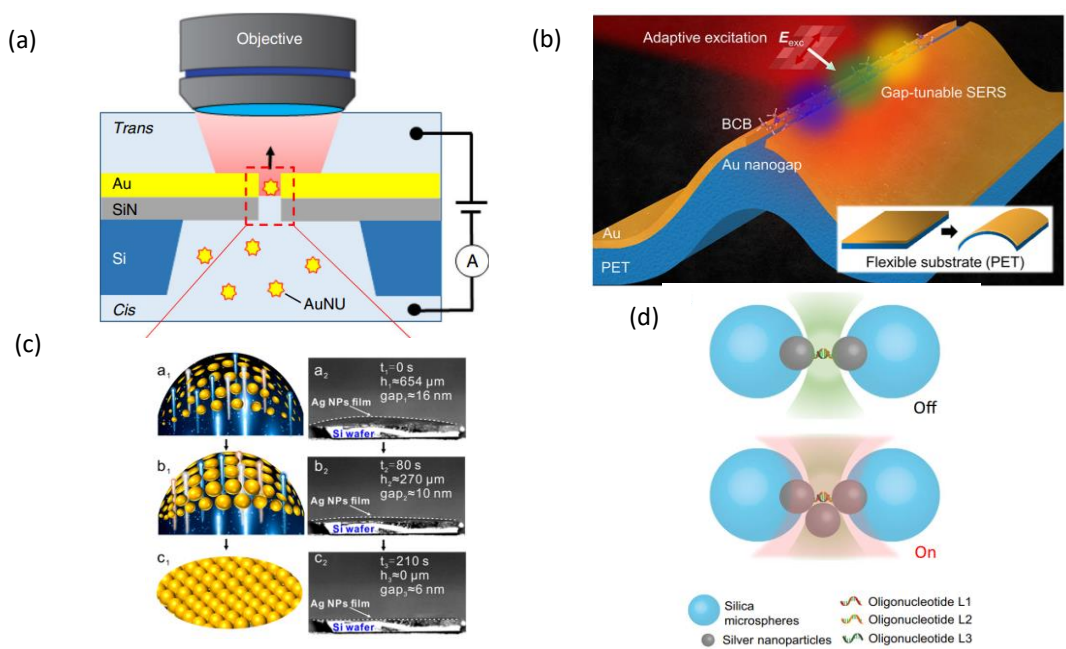

Figure 3 Novel substrates and platforms for dynamic SERS detection. (a) Nanopore to investigate DNA sequences. Reproduced with permission from ref (46). Copyright 2019 Springer Nature. (b) Flexible substrate (PET). Reproduced with permission from ref (53). Copyright 2024 American Chemical Society. (c) AgNPs-decorated Basil-seeds microgels (ABM) platform for dynamic SERS. Reproduced with permission from ref (59). Copyright 2021 Journal of the American Chemical Society. (d) Plasmonic tweezer platform to detect protein conformation. Reproduced with permission from ref (67). Copyright 2023 Springer Nature.

substrates with strong enhancement factors suitable for various applications of dynamic SERS.

In recent years, various nanolithography methods, including electron beam lithography, nanosphere lithography, laser interference lithography, and the emerging technique of displacement Talbot lithography and so on are becoming staple methods for fabricating precise SERS active nanostructures such as nanopores(49) or nanoslits(48). A popular research field focuses on DNA sequencing using nanopore or nanoslit platforms. A typical nanopore is depicted in Figure 3(a), Huang et al. used plasmonic nanopores to directly adsorb molecules onto a gold nanoparticle and then trapping the single nanoparticle in a plasmonic nanohole up to several minutes enabling single molecule detection of both DNA and protein sequence with single nucleotide or amino-acid sensitivity(46, 50). In another recent study, the same authors demonstrated a simple plasmonic nanopore prepared by using nanoporous metal to achieve single molecule DNA sensing(51).

Chang et al. reported the fabrication of Au nanorods on carbon-based nanomaterials as hybrid substrates for high-efficiency dynamic SERS(44). The synergistic effect of the plasmonic Au nanorods and the carbon nanomaterials resulted in enhanced Raman signals and improved stability. Fu et al. developed a high-density nanotip-composed 3D hierarchical

Au/CuS hybrid for sensitive, signal-reproducible, and substrate-recyclable SERS detection(52).

A new direction for plasmonic nanogap fabrication is tunable nanogaps. Moon et al. presented a gap-tunable SERS platform based on a flexible 1D Au nanogap device(53). By engineering the gap width through mechanical bending, they achieve a ~1200 $cm^{-1}$ tunability of the gap plasmon resonance, enabling selective enhancement of different molecular vibrational modes. A typical tunable nanogap structure is illustrated in Figure 3(b). Another method for the manipulation of the gap size is the use of thermos-responsive materials(54-56). For example, Mitomo et al. arranged Au triangular nanoprisms into a 2D array on a thermoresponsive hydrogel surface(54). Furter, the slits between the individual nanoparticles could be shrunk via heating the surface. Such obtained hotspots were used for the detection of single proteins trapped between nanoparticles.

Photoactive materials have also been explored for dynamic SERS substrates. De la Asuncion-Nadal et al. developed photoactive Au@$MoS_2$ micromotors for dynamic SERS sensing, leveraging the photocatalytic properties of $MoS_2$ to enhance the SERS performance(57). On the other hand, Lu et al. utilized metal-organic frameworks (MOFs) involving UiO-66, a photoactive material, as templates to guide the electrochemical lithography of SERS substrates, resulting in high-quality, reproducible plasmonic nanostructures(58).

Most of these studies presented novel substrates with hotspots that have strong enhancement. Moving analyte molecules into hotspots is another direction for enhanced sensitivity. Ge et al. constructed a nanocapillary pumping system for actively capturing target molecules in hotspots formed between nanoparticles(59), Figure 3(c) shows the schematics of initial shrinking stages of AgNPs-decorated Basil-seeds microgels (ABM) and their respective SEM images. In other studies, Chen et al. and Song et al. also developed their respective systems to dynamically capture and detect pesticides in water and fruit peel, respectively(60, 61).

DNA represents another instrument enabling positioning analytes directly into a Raman-active hotspot. Notably, there are different ways DNA can be employed for this purpose. In the simplest approach DNA is simply modified with a specific molecule, typically a fluorophore, to position it into a hotspot. A vivid example of this approach is the creation of a hotspot between core-shell Au@Au NPs with dye-labeled DNA molecules positioned in the interface(62). Likewise, DNA can be used to create a hotspot between two plasmonic nanoparticles, positioning the studied molecule in the hotspot at the same time(63). For example, Lim et al. used this approach to create a Raman sensor for the detection of target DNA strands(64). Once a DNA strand of interest is present in a solution, it hybridizes to DNA strands on gold nanoparticles, creating a dimer and facilitating a measurement of the Raman signal of Cy3 deposited on one of the nanoparticles. Finally, the full potential of the DNA strands is realized with the DNA origami technique. In this method, a 3D DNA-based structures with arbitrary shapes are created via folding of a long, circular plasmid with a set of short DNA strands. Due to the specificity and addressability of the DNA strands, obtained structures create a perfect platform to position both the nanoparticles and studied molecules with a nanometer precision, and thus has been used extensively for the creation of optical nanoantennas, also for SERS(64-66) .

Finally, there is currently a new strategy that can be combined with dynamic SERS, the plasmonic tweezer. The plasmonic tweezer can trap nanoparticles to construct real-time reversible hotspot(67), shown in Figure 3(d). This optical plasmonic tweezer-coupled SERS platform can identify low-populated species among a background of molecules, achieving single-molecule sensitivity, enables real-time monitoring of structural transitions in molecules.

### Data analysis and machine learning algorithms

Data analysis algorithm in dynamic SERS is seeing rapid development due to a) large quantities of spectral data produced by dynamic SERS and b) the development of deep learning methods(68). Machine learning algorithms are also particularly useful for biomolecular detection and classification, because direct analysis of SERS spectra of biological samples is very difficult.

Wang et al. presented a label-free method for characterizing native proteins using SERS and correlation coefficient-directed data analysis(69). They utilized a machine learning approach based on correlation coefficients to differentiate between different proteins and their conformational changes. This method allows for the rapid and accurate identification of proteins without the need for labelling, which is particularly useful for studying proteins in their native state.

In another study with proteins, Almehmadi et al. successfully used principal component analysis (PCA) to differentiate single molecule SERS spectra of albumin-Traut's Reagent (BSA-TR) adduct and the control glycine-Traut's Reagent (glycine-TR) adduct(70).

Dong et al. demonstrated the application of dynamic SERS and support vector machines (SVM) for the detection and direct readout of drugs in human urine samples(71). They collected dynamic SERS spectra from urine samples containing different drugs and used SVM, a supervised machine learning algorithm, to classify and identify the drugs based on their spectral features. This approach enables rapid and sensitive detection of drugs in complex biological matrices, highlighting the potential of combining SERS with machine learning techniques for practical sensing applications.

Machine learning algorithms can also be used on SERS spectra of cells. For example, Fuentes et al. used Raman spectroscopy to measure tumor cell lines and used convolutional neural network (CNN) to accurately classify cell lines as radiosensitive or radioresistant(72).

## Single molecule dynamic SERS and applications

### Monitoring chemical reactions and catalysis

One of the most critical and impactful applications of SM dynamic SERS is to monitor chemical reactions, especially catalysis in real time(73, 74). The landmark work was done by Zee Hwan Kim's group, they demonstrated that by using well-controlled plasmonic junctions and time-resolved SERS, it is possible to directly monitor the metal-catalyzed chemical reactions of individual molecules in real-time, providing a powerful new tool for studying heterogeneous catalytic reactions at the single-molecule level(75). Dynamic SERS allows researchers to identify ephemeral intermediates to gain insights into reaction mechanisms and kinetics.

An et al. employed sub-second time-resolved SERS to investigate the dynamic CO intermediates during electrochemical $CO_2$ reduction on copper electrodes(76). With time-resolved SERS, they were able to reveal a highly dynamic adsorbed CO intermediate, with a characteristic vibration below 2060 cm$^{-1}$ that was correlated with C-C coupling and ethylene production, as shown in Figure 4 (a).

Zoltowski et al. studied the frequency fluctuations of SERS signals during a photocatalytic reaction using a wide-field under sunlight irradiation. Figure 4(b) illustrates the conversion of ROS produced by the MOF through photocatalysis into detectable Raman signals, providing a real-time SERS heatmap of the process.

Dynamic SERS can also be used to study the interaction between molecular scaffolding to the nanoparticles. Ai et al. studied the interaction between cucurbit[7]uril (CB[7]) and gold nanoparticles(GNPs), and were able to observe selective enhancement of normally weak vibrational modes of the CB[7] molecule(78). Figure 4(c) shows a time-resolved SERS trajectory with blinking signals, which was taken ~60 min after adding GNPs applied with a +0.5 V bias. Results show the dynamics of transiently formed CB[7] plasmonic molecular junctions.

In recent years, another important field in chemical research has been the analysis of interface between liquids, liquid/gas or liquid/solid, and dynamic SERS is often suitable for this because of low water background. For example, Yan et al. developed a novel SERS platform capable of monitoring single spectral imaging approach with a frame rate of 100 ms, demonstrating the formation of relevant intermediates and their dependence on nanoparticle aggregation state(77).

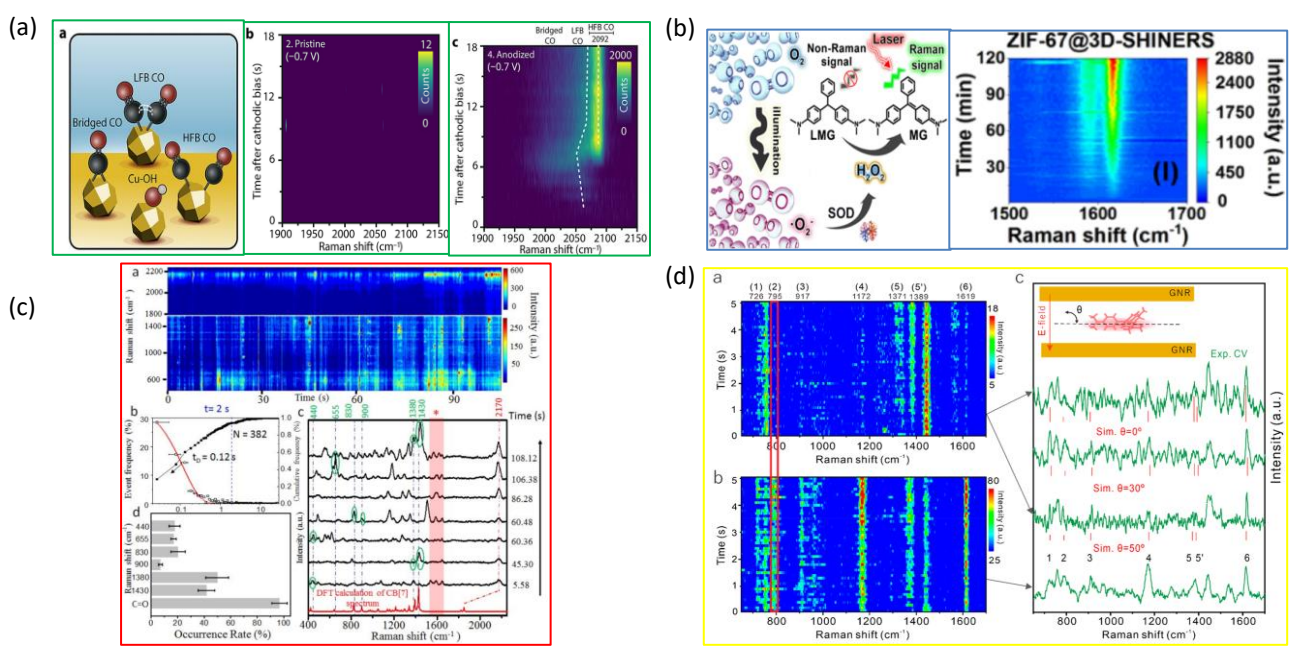

Figure 4 Dynamic SERS researching on (a) Intermediates during chemical reduction, with spectral heatmap from SERS measurements showing the dynamic behavior of corresponding reduced steps. Reproduced with permission from ref (76). Copyright 2021 Wiley-VCH GmbH. (b) Photocatalysis. Reproduced with permission from ref (22). Copyright 2022 American Chemical Society. (c) Charge transfer between molecular scaffolding to the nanoparticles. Reproduced with permission from ref (78). Copyright 2020 The Royal Society of Chemistry. (d) Single molecular motion at the gas/liquid interface. Reproduced with permission from ref (79). Copyright 2023 American Chemical Society.

One of the frontiers for dynamic monitoring of chemical reactions is photocatalysis. Chen et al. developed a de-Au@mTiO$_2$ nanocage reactor, with dense shareable hot spots for in-situ and dynamic SERS tracking of photocatalysis(45). In another study, Cheng et al. demonstrated in-situ monitoring of dynamic photocatalysis using MOFs decorated with shell-isolated nanoparticles (SHINs)(22). Using dynamic SERS, Cheng et al. compared between different MOF photocatalysts, including ZIF-67, ZIF-8, and UIO-66 and recorded real-time spectroscopic evidence of reactive oxygen species formation molecules in real-time at the gas/liquid interface for extended periods**(79)**, as illustrated in Figure 4(d). By combining millisecond dynamic SERS with DFT calculations, the authors captured the orientation of individual CV molecules moving within the sub-nanometer space, accomplishing real time molecular orientation monitoring.

**Single molecule detection and sensing**

Dynamic detection and sensing of single molecules is also an important avenue of SM dynamic SERS. Changes in biomarkers can often indicate diseases, drug effects or health changes, and the ability to detect such changes in real time could help future diagnosis of diseases or drug development immensely.

In 2020, Zong et al. was able to detect single R6G molecules probe single molecules transporting through the nanopore using dynamic SERS detection. For small rhodamine 6G (R6G) molecules, the SERS spectra varied based on the orientation of the molecules during transport. New peaks and peak shifts were observed, likely due to charge transfer interactions between R6G and the gold surface in different orientations, which shows in Figure 5(a), (b), and (c). For single hemoglobin (Hb) proteins,

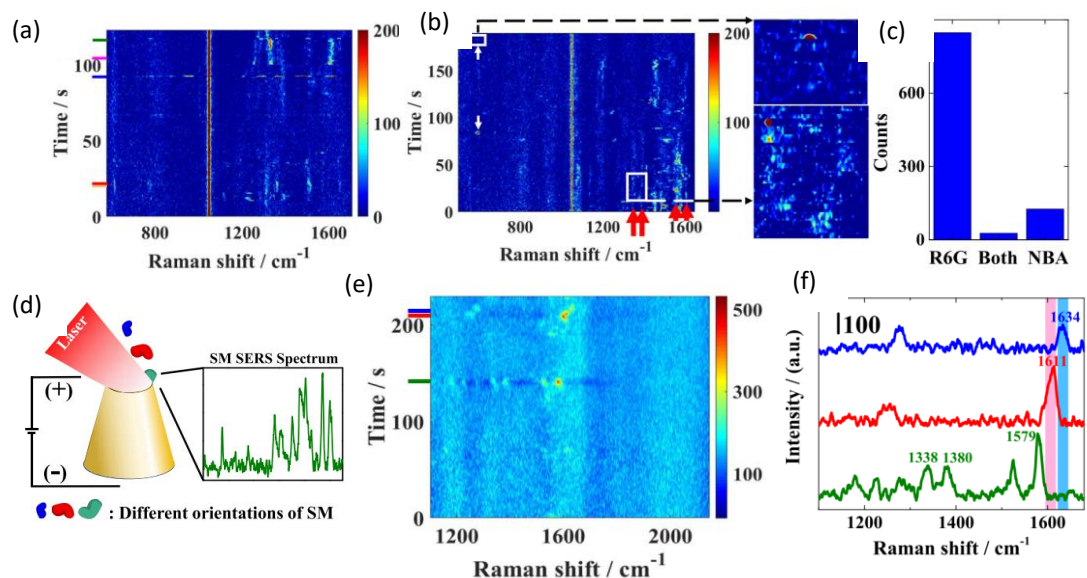

Figure 5 (a-b) Contour graphs of Raman spectra of $10^{-9}$ M R6G, the mixture of $10^{-9}$ M R6G and $10^{-9}$ M NBA under −300 mV with a 0.3 s exposure time respectively; (c) Bi-analyte SERS distribution of R6G and NBA; (d) Conjoint SERS−plasmonic nanopore setup; (e) Contour Raman spectra of $10^{-9}$ M Hb in a 1 M LiNO3 (pH 6.6) solution under −300 mV with a 2 s exposure time. The SERS spectra shown in (f) are labeled by short lines with corresponding colors. (f) The SERS spectra extracted from (e) at the time 114, 209, and 211 s, respectively. Reproduced with permission from ref (82). Copyright 2021 American Chemical Society.

with 50 ms temporal resolution, and captured the transient neutral R6G0 state of R6G on AgNPs(24). Li et al. developed a digital nanopillar SERS platform for ultrasensitive multiplex cytokine analysis(80). This platform can be used to predict and monitor immune-related adverse events (irAEs) in cancer immunotherapy. Their assay enables single cytokine counting down to attomolar levels, and the platform demonstrated its capability to longitudinally monitor cytokine levels and predict severe immune-related adverse event risk in a pilot cohort of melanoma patients.

Dynamic SERS detection can also be used in environmental monitoring; Guo et al. presented a rotary micromotor-sensor system that accelerates the enrichment and detection of DNA molecules. The microsensors are composed of diatom frustules adorned with AgNPs, and can reach an enhancement factor of $10^9$ to $10^{10}$. This system is used to monitor the presence of single DNA molecules in water samples(81).

Nanopore are also used for the detection and sensing of a single molecule or of its orientation, electronic state, and oxygenation states. Using a sub-10 nm conical gold nanopore, Zhou et al. was able to probe multidimensional structural information of single molecules while they were translocated through the nanopores(82). As shown in Figure 5(d), they fabricated a sub-10 nm conical gold nanopore at the tip of a quartz nanopipette. This plasmonic nanopore allowed them to as shown in Figure 5(e) and (f), the SERS spectra could distinguish the orientation of the porphyrin ring (vertical vs parallel) as well as the oxygenated vs deoxygenated states of Hb as it passed through the nanopore.

Described methods enable a single molecule detection and sensing, however they lack a control over the position of the analyte in space. Detection is therefore based solely on measurements of molecules randomly passing through the hotspot area or deposited at different locations on the Raman-active surface. First, it extends the measuring time until the passing molecule is observed. Secondly, this can result in large differences in signal intensity for molecules oriented in different hotspot areas(83). Finally, if a molecule passes through a hotspot or easily detaches from a surface, the available measurement time may be too short to observe the dynamic processes of the molecules under investigation.

This problem can be solved using the already introduced DNA origami technique. As explained before, with this method, it is not only easy to create a plasmonic hotspot, but also to precisely place the molecule of interest within it. For this reason, the DNA origami technique has been extensively employed for single-molecule measurements(66). Vibrant example of the use of DNA origami for the SM measurements is the detection of a single proteins(84-86). For example, the detection of a single streptavidin and thrombin protein was realized by placing them between gold nanowires using biotin or an aptamer and measuring the Raman signal over time(85).

however, molecules or structures, such as adenine or protein tertiary structures, can be studied using SERS as well.

Huang's group at HKUST pioneered the work of combining SERS with optical tweezers for single biomolecular assays. In

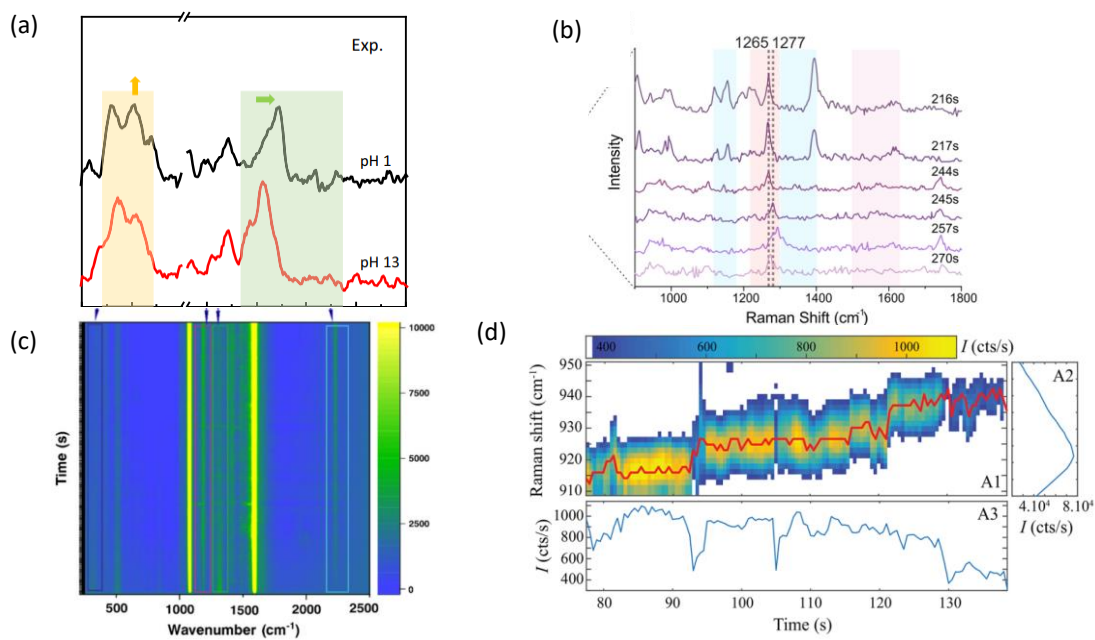

Figure 6 Dynamic interaction or conformation of molecule. (a) SERS spectra of Tyr at pH 1.0 (black) and pH 13.0 (red), respectively. Reproduced with permission from ref (67). Copyright 2023 Springer Nature. (b) SERS spectra showing different time points from a single DONA measurement, highlighting the shift of the Amide III band (blue zones highlight heme unit peaks, the red zone indicates peptide backbone peaks, and the purple zone shows spectral contributions from both). Reproduced with permission from ref (87). Copyright 2024 Springer Nature. (c) Time-series SERS map recorded for 100 s under 633 nm laser excitation showing the transformation of NMSB molecule. Reproduced with permission from ref (73). Copyright 2024 American Chemical Society. (d) Surface enhanced Raman chromatograms of three specific Raman lines of Trp in the single molecule regime. Reproduced with permission from ref (90). Copyright 2019 American Chemical Society.

In another example single cytochrome C proteins were monitored over time and spectral shifts of the Amide III band indicated conformational changes of the protein within the hot spot(87). Furthermore, the catalytic cycle of a single horseradish peroxidase (HRP) enzyme was monitored over time using dynamic SERS and DNA origami nanostructures(88).

The ability to detect and characterize single molecules using nanopores and DNA origami technique opens new possibilities for studying molecular orientation and sequencing at the single-molecule level. With the help of trapping tools such as nanopores or plasmonic traps, studying biomolecular interactions and conformational dynamics are becoming a frontier for dynamic SERS. These tools also enable researchers to perform dynamic SERS sequencing of nucleic acids and proteins. In fact, Zhou et al's work(82) laid the foundations for their future work of protein sequencing. These applications will be discussed in the following sections.

**Bio-molecular interactions and conformational dynamics**

As previously mentioned, SM dynamic SERS has significant applications for investigating biomolecular interactions. Plasmonic, molecular binding or nanopores are often used to trap single biomolecule for SM dynamic SERS. Many of the studies on conformational dynamics are applied to proteins, peptides or even individual amino acids while some are done on DNA or RNA as well. Plasmonic tweezers are often combined with fluorescence to study bio-molecular interaction(89);

2023, they demonstrated the use of an efficient optical plasmonic tweezer-controlled SERS platform for the characterization of pH-dependent amylin species in aqueous milieus(67). As shown in Figure 6(a), by optically trapping a silver nanoparticle at the plasmonic junction between two AgNP-coated silica microbeads, they created a dynamic nanocavity that acquired Raman spectra, which provides information on the distribution of protein tertiary structures in real time. This allowed them to characterize the pH-dependent conformational changes of amylin in real time. Mostafa et al. also used AuNP dimers to study protein conformational changes, and observed the oxidation and conformational change processes of cytochrome C(87). In addition, they also studied the reactivity of N-methyl-4-sulfanylbenzamide (NMSB) at nanocavities of gold and silver nanoparticle aggregates under plasmonic excitation to gain understanding of the respective reaction mechanism(73). The relevant results are shown in Figure 6(b) and (c), respectively.

Wang et al. developed a method to characterize protein "native-ness" using dynamic SERS(69). With 3D physiological hotspots formed with iodide-modified gold nanoparticles, they were able to quantify the degree of protein denaturation caused by various ions. This suggests that using dynamic SERS to observe changes in protein conformation induced by external factors is promising.

Leray et al. also used dynamic SERS to resolve conformational changes and charge transfer in the amino acid tryptophan (Trp) at the single-molecule level(90). As shown in Figure 6(d), they were able to discern the conformational changes by analyzing the auto- and cross-correlation functions of specific Raman bands.

For DNA conformational changes, Feng et al. designed a SERS-based molecular ruler to directly monitor the conformational changes of single DNA aptamers in real-time(91). They designed a core-satellite plasmonic

Sugano et al. also demonstrated the dynamic SERS detection of a single DNA oligomer using a single AuNP dimer(93). By controlling the position and orientation of the AuNP dimer, they ensured a single hotspot in the laser spot. The Raman peaks of DNA components, including nucleobases and the backbone, were dynamically observed over time. The protein structure with time-dependent SERS spectra is shown in Figure 7(b).

Nanopores can also be used to study and sequence proteins. Zhou et al used a plasmonic nanopore for protein sequencing(47). As illustrated in Figure 7(c), by applying high nanostructure where the ~20 nm AuNP satellites are connected to the ~100 nm core AuNP via a single DNA aptamer. Upon binding to its target, the aptamer undergoes a conformational change, bringing the satellite closer to the core. This results in a stepwise enhancement in the SERS signal intensity due to increased plasmonic coupling, and the authors were able to track the dynamic folding processes in real time.

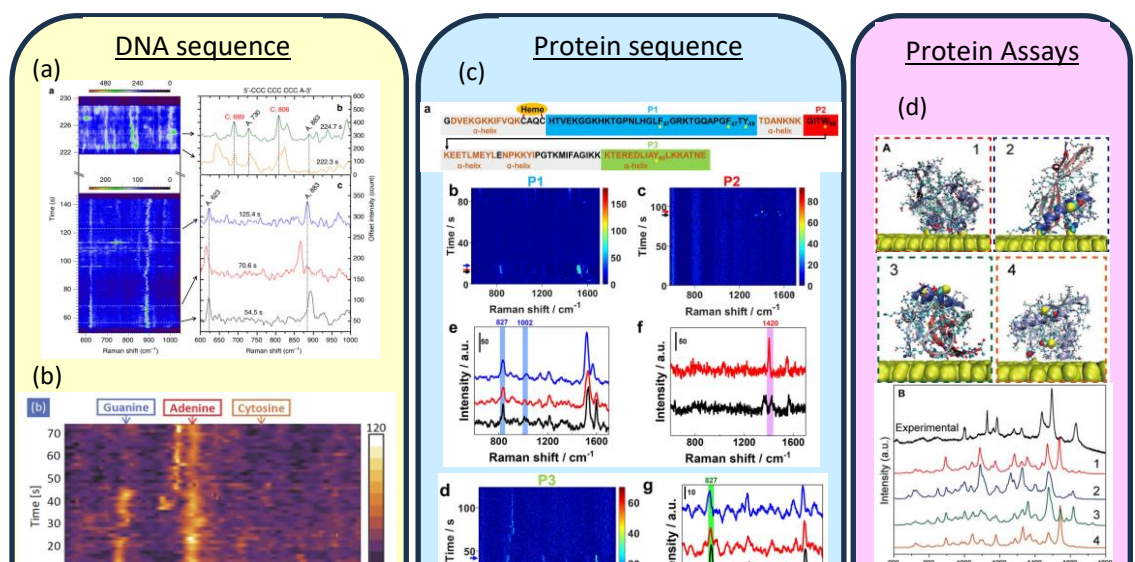

Figure 7 Sequences or structural assays detection with time-dependent SERS spectra. (a) DNA sequence detection by nanopore. Reproduced with permission of ref (46). Copyright 2019 Springer Nature. (b) DNA sequence research from a gold nanoparticle dimer. Reproduced with permission of ref (93). Copyright 2022 Optica Publishing Group. (c) Investigated on protein sequence by a plasmonic structure. Reproduced with permission of ref (47). Copyright 2023 American Chemical Society. (d) Rapidly determining the 3D structure of protein assays. Reproduced with permission of ref (94). Copyright 2023 American Association for the Advancement of Science.

**Nucleic acid and protein sequencing and structural assays**

Dynamic SERS can be combined with nanopores and nanoslit structures to study nucleic acids. By using a nanopore-based SERS platform, Huang et al. was able to detect the translocation of individual nucleic acid molecules through the nanopore in real-time, as shown in Figure 7(a), while simultaneously obtaining sequence-specific SERS signals(46, 50, 92). To achieve it, they used a plasmonic platform that enable the electro-plasmonic trapping of individual Au nanoparticles. Thanks to this effect it was possible to generate a nanocavity between the particle and the nanopore, so obtaining a giant field localized in few nm. Other groups, such as Chen et al, also achieved dynamic detection of single nucleobase using nanopore or nanoslit based SERS platforms(48).

bias voltages, single cytochrome c proteins were unfolded and sequentially translocated through a gold nanopore. Characteristic SERS spectra of specific amino acids were recorded as different protein segments passed through the SERS hotspot, making it possible to use dynamic SERS to sequence proteins.

Advancements in plasmonic nanocavities not only enables SERS based DNA or protein sequencing, they also accelerate the development of SERS based protein structural assays. As seen in Figure 7(d), In 2023, Ma et al. used a nanoparticle-on-mirror (NPoM) setup with a deciphering strategy called SPARC to accurately reconstruct protein structures based on SERS signals in seconds(94). The development of such methods could lead to the development of dynamic structural analysis of interfacial proteins in their native environments.

**Discussion, challenges and future directions**

One of the main challenges of single molecule dynamic SERS is its repeatability and reliability(95). Many factors could contribute to the reliability and repeatability of dynamic SERS, such as substrate design, sample preparation, control of measurement conditions, and data analysis. Reproducibility and uniformity of SERS substrates is one of the problems that contribute to the less-than-ideal repeatability of dynamic SERS(96, 97). Further research into standardized protocols and robust data analysis methods may help overcome these limitations.

Complex environments may also hinder the application of dynamic SERS in-situ, and dynamic SERS measurements in-vivo are even more difficult. While there are some studies that achieved in-situ single molecule SERS, such as the works of Li et al. and Yang et al., studies about in-situ single molecule dynamic SERS are rare(98, 99).

While there are already studies that combined dynamic SERS to other cutting-edge technologies such as plasmonic trapping or nanopores, other frontier plasmonic technologies may also be combined with dynamic SERS to great effects.

Metasurfaces, for example, could be combined with dynamic SERS(100). Martins et al. combined metasurfaces with ultrafast low field-of-view (FoV) deflectors to achieve high frame rates (kHz) and a large FoV for LiDAR, which has potential use in dynamic SERS.

The development of UV SERS also holds promise(101). UV SERS presents several promising advantages, such as increased scattering intensity, higher spatial resolution, resonance enhancement for organic, biological, and semiconductor analytes. However, further development of enhancement substrates is required for single molecule dynamic SERS in UV ranges; Currently aluminum nanostructures are the best-performing substrates, while future substrates may push the sensitivity to single molecule level(102-104).

The future of dynamic SERS is promising, with ongoing advancements in SERS substrates, instrumentation, and data analysis techniques. Key future research fields are improved sensitivity, temporal resolution and data process techniques. Combining dynamic SERS with developing plasmonic nanotechnology, including nanopores and plasmonic tweezers may also open new avenues for single molecule assays. Finally, future research in stability and repeatability may improve the application of dynamic SERS both in-situ and potentially in-vivo.

## Conflicts of interest

There are no conflicts to declare.

## Acknowledgements

The authors acknowledge financial support from NATIONAL NATURAL SCIENCE FOUNDATION OF CHINA, Grant No. 22202167 and NATIONAL KEY RESEARCH AND DEVELOPMENT PROJECT OF CHINA, Grant No. 2023YFF0613603; we also thanks the European Union under the Horizon-MSCA-DN-2022: DYNAMO, grant Agreement 101072818.

## Notes and references